\documentstyle[12pt]{article}

\author{Hao Chen\\
Department of Mathematics\\
Zhongshan University\\
Guangzhou,Guangdong 510275\\
People's Republic of China\\
and\\
Department of Computer Science\\
National University of Singapore\\
Singapore 117543\\
Republic of Singapore}
\title{Schmidt Number of Pure States in Bipartite Quantum Systems as an Algebraic-geometric Invariant}
\date{August,2001}

\begin{document}

\maketitle
\begin{abstract}

In [1] and [2] we defined new invariants of both bipartite and multipartite quantum systems under local unitary transformations via algebraic-geometry of determinantal varieties, and gave a new separability criterion based on our new invariants. The purpose of this note is to show how Schmidt number of pure states in bipartite systems, a classical invariant, can be read out from algebraic-geometric invariants. 
\end{abstract}

In [1] and [2], we introduced algebraic sets (ie., the zero locus of several multi-variable homogeneous polynomials, see [4]) in the complex projective spaces and the products of complex projective spaces for mixed states in a bipartite and multipartite quantum systems with the following two properties:\\

1) When we apply local unitary operations to the mixed state, the corresponding  algebraic sets are changed by a linear transformation, and thus these invariants can be used to distinguish inequivalent mixed states under local unitary  operations;\\

2)The algebraic sets are linear (the union of some linear subspaces) if the mixed state is separable, and thus we give a new separability criterion.\\

Actually as indicated in [1] and [2], if the standard Euclid metric of the complex space (resp. Fubini-Study metric of the projective complex space) is used, the metric properties of the algebraic sets are also preserved when local  unitary operations are applied to the mixed state. \\

Though the invariants defined in [1] and [2] cannot be used for very low rank mixed states, for example, the invariants there for pure states are always the whole space. However we modify the definition there and thus introduce more algebraic-geometric invariants of the mixed states in this note. The main example is Theorem 4 which shows that Schmidt number of pure states is actually an algebraic-geometric invariant.\\

Let $H=C_{A}^m \otimes C_{B}^n$ and the standard orthogonal base is $\{|ij>\}$, where, $i=1,...,m$ and $j=1,...,n$, and $\rho$ is a mixed state on $H$. We represent the matrix of $\rho$ in the base $\{|11>,...|1n>,...,|m1>,...,|mn>\}$, and consider $\rho$ as a blocked matrix $\rho=(\rho_{ij})_{1 \leq i \leq m, 1 \leq j \leq m}$ with each block $\rho_{ij}$ a $n \times n$ matrix corresponding to the $|i1>,...,|in>$ rows and the $|j1>,...,|jn>$ columns. We recall the definition of algebraic-geometric invariants in [1].\\

{\bf Definition 1.}{\em  We define 

$$
\begin{array}{ccccc}
V_{A}(\rho)=\{(r_1,...,r_m)\in CP^{m-1}:det( \Sigma_{i,j}r_ir_j^{*} \rho_{ij})=0\}
\end{array}
$$
Similarly $V_{B}(\rho) \subseteq C^n$ can be defined. Here * means the conjugate of complex numbers and det means the determinant.}\\

{\bf Example 1.} Let $\rho=\frac{1}{mn}I_{mn}$, the maximally mixed state, we easily have $V_{A}(\rho)=\{(r_1,...,r_n): det(\Sigma r_i r_i^{*}I_n)=0\}=\emptyset$. This is a trivial example.\\

It is thus natural we can define the following generalized invariants.\\

{\bf Definition 2.}{\em  We define 

$$
\begin{array}{ccccc}
V_{A}^k(\rho)=\{(r_1,...,r_m)\in CP^{m-1}:rank( \Sigma_{i,j}r_ir_j^{*} \rho_{ij}) \leq k\}
\end{array}
$$
, for $k=0,1,...,n-1$. Similarly $V_{B}^k(\rho) \subseteq CP^{n-1}$ can be defined.}\\

It is clear that $V_A(\rho)=V_A^{n-1}(\rho)$ here, thus the invariants introduced here are natural generalization of invariants in [1].\\

Similarly as in [1] we have the following results.\\

{\bf Theorem 1.}{\em  Let $T=U_{A} \otimes U_{B}$, where $U_{A}$ and $U_{B}$ are the local operations (ie., unitary linear transformation) on $C_{A}^m$ and $C_{B}^n$ rescpectively. Then $V_{A}^k(T(\rho))=U_{A}^{-1}(V_{A}^k(\rho))$, that is $V_{A}^k(\rho)$ (resp. $V_{B}^k(\rho)$) is a ``invariant'' upto a linear transformation of $CP^{m-1}$ of the mixed state $\rho$ under local unitary operation.}\\

{\bf Proof.} Let $U_{A}=(u_{ij}^{A})_{1 \leq i \leq m,1 \leq j \leq m}$, and $U_{B}=(u_{ij}^{B})_{1 \leq i \leq n, 1 \leq j \leq n}$, be the matrix in the standard orthogonal bases. Then the matrix of $T(\rho)$ under the standard orthogonal base $\{|ij>\}$, where $ 1 \leq i \leq m, 1 \leq j \leq n$, is $T(\rho)=(\Sigma_{l,k} u_{il}^{A} U_{B}(\rho_{lk})(U_{B}^{*})^{ \tau}(u_{jk}^{A})^{*})_{1\leq i \leq m, 1 \leq j \leq m}$. Hence \\

$$
\begin{array}{cccccccc}
V_{A}^k(T(\rho))=\{(r_1,...,r_m): rank(\Sigma _{l,k} (\Sigma_i r_i u_{il}^{A})(\Sigma_i r_i u_{il}^{A})^{*}U_{B}(\rho_{lk})(U_{B}^{*})^{\tau}) \leq k\}

\end{array}
(1)
$$

We set $r_l'=\Sigma_i r_i u_{il}^{A}$ for $l=1,...,m$.Then\\

$$
\begin{array}{ccccc} 
\Sigma _{l,k} (\Sigma_i r_i u_{il}^{A})(\Sigma_i r_i u_{il}^{A})^{*}U_{B}(\rho_{lk})(U_{B}^{*})^{\tau}\\
=U_{B}(\Sigma_{lk} r_l' (r_k')^{*} (\rho_{lk})(U_{B}^{*} )^{\tau}

\end{array}
(2)
$$

Thus\\
$$
\begin{array}{ccccccc}
V_A^k (T(\rho))=\{(r_1,...,r_m):rank(\Sigma_{lk} r_l' (r_k')^{*} (\rho_{lk})) \leq k\}

\end{array}
(3)
$$

and our conclusion follows. \\

{\bf Remark 1.} Since $U_{A}^{-1}$ certainly preserve the standard Euclid metric $\Sigma r_ir_i^{*}$ of $C^m$ (resp. Fubini-Study metric of $CP^{m-1}$), we know that all metric properties of $V_{A}(\rho)$ are preserved when the local operations are applied to the mixed state $\rho$.\\

In the following statement, the term ``algebraic set `` means the set of  zeros of a system of $m$ variable polynomials.(see [18]).\\

{\bf Theorem 2.} {\em $V_{A}^k(\rho)$ (resp. $V_{B}^k(\rho)$) is an algebraic set in $CP^{m-1}$ (resp.$CP^{n-1}$).}\\

{\bf Theorem 3.} {\em If $\rho$ is a separable mixed state, $V_{A}^k(\rho)$ (resp. $V_{B}^k(\rho)$) is a linear subset of $CP^{m-1}$ (resp.$CP^{n-1}$),ie., it is the union of the linear subspaces.}\\

For the purpose to prove Theorem 2 and 3 we need the following lemmas.\\

{\bf Lemma 1.} {\em We take the orthogonal base $\{e_1,...,e_h\}$ of a $h$ dimension Hilbert space $H$. Let $\rho=\Sigma_{l=1}^{t} p_l P_{v_l}$, where $v_l$ is unit vector in $H$ for $l=1,...,t$ ,   $v_l=\Sigma_{k=1}^{h} a_{kl} e_k$ , $A=(a_{kl})_{1\leq k \leq h, 1 \leq l \leq t}$ is the $h \times t$ matrix and $P_{v_l}$ is the projection to the vector $v_l$. Then the matrix of $\rho$ with the base $\{e_1,...,e_h\}$ is $AP(A^{*})^{\tau}$, where $P$ is the diagonal matrix with diagonal entries $p_1,...,p_h$.}\\

{\bf Proof}. We note that the matrix of $P_{v_l}$ with the base is $\alpha (\alpha^{*})^{\tau}$ where $\alpha=(a_{1l},...,a_{hl})^{\tau}$ is just the representation vector of $v_l$ with the base. The conclusion follows immediately.\\

The following conclusion is a direct matrix computation from Lemma 1 or see [5],[6]\\

{\bf Corollary 1.} {\em Suppose $p_i>0$, then the image of $\rho$ is the linear span of vectors $v_1,...,v_t$.}\\

Now let $H$ be the $C_{A}^m \otimes C_{B}^n$ ,$\{e_1,...,e_{mn}\}$ be the standard orthogonal base $\{|11>,...,|1n>,...,|m1>,...,|mn> \}$ and $\rho= \Sigma_{l=1}^{t} p_l P_{v_l}$ be a mixed state on $H$. We may consider the $mn\times t$ matrix $A$ as a $m\times 1$ blocked matrix with each block $A_w$, where $w=1,...,m$, a $n\times t$ matrix corresponding to $\{|w1>,...,|wn>\}$. Then it is easy to see $\rho_{ij}=A_iP(A_j^{*})^{\tau}$, where $i=1,...m,j=1,...,m$. Thus\\

$$
\begin{array}{cccccc}
\Sigma r_ir_j^{*} \rho_{ij}=(\Sigma r_i A_i)P(\Sigma r_i^{*} A_i^{*})^{\tau}
\end{array}
(4)
$$

{\bf Lemma 2.} {\em $\Sigma r_ir_j^{*} \rho_{ij}$ is a (semi) positive $n\times n$ matrix. Its rank equal to the rank of $(\Sigma r_i A_i)$.}\\

{\bf Proof.} The first conclusion is clear. The matrix is of rank $\leq k$ if and only if there exist $n-k$ linear indepandent vectors $c^j=(c_1^j,...,c_n^j)$ ,for $j=1,...,n-k$ with the property.\\

$$
\begin{array}{cccccccccc}
c^j(\Sigma r_ir_j^{*} \rho_{ij})(c^{j*})^{\tau}=\\
(\Sigma r_i c^jA_i)P(\Sigma r_i^{*} c^{j*}A_i^{*})^{\tau}=0
\end{array}
(5)
$$

Since $P$ is a strictly positive definite matrix,our conclusion follows immediately.\\

{\bf Proof of Theorem 2.} From Lemma 2 , we know that $V_{A}^k(\rho)$ is the zero locus of all $(k+1)\times (k+1)$ submatrices of $(\Sigma r_i A_i)$ in $CP^{m-1}$.\\

Now suppose that the mixed state $\rho$ is separable,ie, there are unit product vectors $a_1 \otimes b_1,....,a_s\otimes b_s$ such that $\rho=\Sigma_{l=1}^{s}q_l P_{a_l \otimes b_l}$ , where $q_1,...q_s$ are positive real numbers. Let $a_u=a_u^1 |1>+...+a_u^m |m>,b_u= b_u^1 |1>+...+b_u^n|n>$ for $u=1,...,s$. Hence the vector representation of $a_u \otimes b_u$ with the standard base is $a_u \otimes b_u= \Sigma_{ij} a_u^ib_u^j |ij>$. Consider the corresponding $mn \times s$ matrix $C$ of $a_1 \otimes b_1,...,a_s \otimes b_s$ as in Lemma 1, we have $\rho=CQ(C^{*})^{\tau}$, where $Q$ is diagonal matrix with diagonal entries $q_1,...,q_s$. As before we consider $C$ as $m\times 1$ blocked matrix with blocks $C_w$, $w=1,...m$. Here $C_w$ is a $n \times s$ matrix of the form $C_w=(a_j^{w}b_j^i)_{1\leq i \leq n, 1 \leq j \leq s}=BT_w$ , where $B=(b_j^i)_{1\leq i \leq n, 1\leq j\leq s}$ is a $n \times s$ matrix and $T_w$ is a diagonal matrix with diagonal entries $a_1^{w},...,a_{s}^{w}$. Thus from Lemma 1, we have $\rho_{ij}=C_i Q (C_j^{*})^{\tau}=B(T_i Q (T_j^{*})^{\tau})(B^{*})^{\tau}=BT_{ij}(B^{*})^{\tau}$, where $T_{ij}$ is a diagonal matrix with diagonal entries $q_1 a_1^i(a_1^j)^{*},...,q_s a_s^i(a_s^{j})^{*}$.\\

{\bf Proof of Theorem 3.} As in the proof of Theorem 2, we have\\

$$
\begin{array}{ccccccc}
\Sigma r_ir_j^{*} \rho_{ij}=\Sigma r_i r_j^{*}BT_{ij}(B^{*})^{\tau}\\
B(\Sigma r_ir_j^{*}T_{ij})(B^{*})^{\tau}
\end{array}
(6)
$$

Here we note  $\Sigma r_ir_j^{*}T_{ij}$ is a diagonal matrix with diagonal entries\\
$ q_1(\Sigma r_ia_1^{i})(\Sigma r_ia_1^{i})^{*},...,q_s(\Sigma r_ia_s^{i})(\Sigma r_ia_s^{i})^{*}$.Thus $\Sigma r_ir_j^{*} \rho_{ij}=BGQ(G^{*})^{\tau}(B^{*})^{\tau}$, where $G$ is a diagonal matrix with diagonal entries $ \Sigma r_ia_1^{i},...,\Sigma r_ia_s^{i}$. Because $Q$ is a strictly positive definite matrix, from lemma 2 we know that  $rank(\Sigma r_ir_j^{*} \rho_{ij}) \leq k$ if and only if $rank(BG) \leq k$. Note that $BG$ is just the multiplication of $s$ diagonal entries of $G$ (which is linear forms of $r_1,...,r_m$) on the $s$ columns of $B$,  thus the determinants of all $(k+1) \times (k+1)$ submatrices of $BG$ are the multiplications of a constant (possibly zero) and $(k+1)$ linear forms of $r_1,...,r_m$. Thus the conlusion is proved.\\

From Theorem 3 here, actually a stronger separability criterion than in [1] is given. We believe that Theorem 3 in [1] and here are good tools to prove a mixed state is entangled.\\

However as showed in the following example, $V_A^0(\rho)$ for the rank 1 mixed state (ie., pure states) are always linear. In this case our separability criterion cannot work here. Interestingly, Schmidt numbers of pure states $\rho$ can be read out from $V_A^0(\rho)$, and thus actually our invariant still work for the entanglement of even pure states.\\

For the algebraic set $V \subseteq CP^{m-1}$ , it is a standard fact ([4]) in algebraic geometry that $V$ is the sum of its irreducible components, ie., $V=V_1 \cup \cdots \cup V_t$. We define $dim(V)=max \{dim(V_i):i=1,...,t\}$.\\

{\bf Example 2.} Let $H=C_A^m \otimes C_B^n$ ,where, $ m \leq n$, be a bipartite quantum system and $\rho=P_v$, where $v=\Sigma_{ij} a_{ij} |ij>$, be a pure state. Then we can see that $V_A^0(\rho)$ is the locus of the condition that $rank(R) \leq 0$, ie., $R=0$.\\

$$
R=
\left(
\begin{array}{cccccccc}
a_{11}r_1+a_{21}r_2+...+a_{m1}r_m\\
a_{12}r_1+a_{22}r_2+...+a_{m2}r_m\\
......\\
a_{1n}r_1+a_{2n}r_2+...+a_{mn}r_m
\end{array}
\right)
(7)
$$

{\bf Proposition 1.} {\em If $dim(V_A^0(\rho))=m-2$, $v$ is a separable pure state.}\\

{\bf Proof.} In this case $V_A^0(\rho)$ is just the intersection of $n$ hyperplanes in $CP^{m-1}$. If $dim(V_A^0(\rho)=m-2$, the $n$ linear forms in the matrix $R$ has to all linear dependent on one of them. Without loss of generality, suppose $a_{1j}r_1+a_{2j}r_2+...+a_{mj}r_m=c_j(a_{11}r_1+a_{21}r_2+...+a_{m1}r_m)$ for $j=2,...,n$. Thus $v=(\Sigma_i a_{i1}|i>) \otimes (|1>+c_2|2>+...+c_n|n>)$, a separable pure state.\\

Actually from Theorem 1 about the invariance of $V_A^0(\rho)$, we can compute it from its Hilbert-Schmidt decomposition (see [3]) $v= \Sigma_{i=1}^d a_i e_i \otimes e'_i$, where $e_1,...,e_m$ (resp., $e'_1,...,e'_n$) is a orthogonal base of $C_A^m$ (resp. $C_B^n$). In this representation of $v$, $R=\Sigma_i r_i A_i$ in the prrof of Theorem 2,3 above becomes the form $R=(a_1r_1,...,a_d r_d,0,...,0)^{\tau}$.\\

{\bf Theorem 4.} {\em For the pure state $\rho=P_v$, $d=m$ if and only if $V_A^0(\rho)=\emptyset$ and $d=m-1-dim(V_A^0(\rho))$ if $d \leq m-1$.}\\

{\bf Proof.}. From the above representation of $R$, we have $dim(V_A^0(\rho))=m-1-d$, thus the conclusion follows immediately.\\

From Theorem 4 here, we show that Schmidt number of the pure states, a classical invariant of the pure states in bipartite quantum systems, is actually an algebraic-geometric invariant. At the present stage, we are not clear about the cases of other different generalizations of Schmidt numbers for the mixed states ([7],[8],[9]).\\

The author acknowledges the support from NNSF China, Information Science Division, grant 69972049.\\

e-mail: dcschenh@nus.edu.sg\\

\begin{center}
REFERENCES
\end{center}

1.Hao Chen, New invariants and separability criterion of the mixed states: bipartite case, preprint, July, 2001,quantu-ph/0107111\\

2.Hao Chen, New invariants and separability criterion of the mixed states: multipartite case, preprint, July, 2001,quant-ph/0107112\\

3.J.Preskill, Physics 229: Advanced mathematical methods of Physics--Quantum Computation and Information (California Institute of Technology, Pasadena,CA,1998), http://www.theory.caltech.edu/people/preskill/ph229/.\\

4.J.Harris, Algebraic geometry, A first Course, Gradute Texts in Mathematics, 133, Springer-Verlag, 1992,
  especially its Lecture 9  ``Determinantal Varieties''\\

5.A.Peres, Phys.Rev.Lett.77,1413(1996)\\

6.P.Horodecki, Phys.Lett.A 232 333(1997)\\

7.B.Terhal and P.Horodecki, Phys.Rev.A61,040301 (2000)\\

8.A.SanperA, D.Bruss and M.Lewenstein, Phys.Rev.A63, R050301 (2001)\\

9.J.Eisert and H.J.Briegel,Phys.Rev.,A64, 022306(2001) 

\end{document}